\documentclass[%
 aip,
 amsmath,amssymb,
 reprint,%
]{revtex4-1}

\usepackage{graphicx}
\usepackage{dcolumn}
\usepackage{bm}

\usepackage[utf8]{inputenc}
\usepackage[T1]{fontenc}
\usepackage{mathptmx}
\usepackage{etoolbox}

\makeatletter
\def\@email#1#2{%
 \endgroup
 \patchcmd{\titleblock@produce}
  {\frontmatter@RRAPformat}
  {\frontmatter@RRAPformat{\produce@RRAP{*#1\href{mailto:#2}{#2}}}\frontmatter@RRAPformat}
  {}{}
}%
\makeatother

\usepackage{xcolor}

\DeclareUnicodeCharacter{2212}{-}

\begin{document}

\preprint{AIP/123-QED}

\title[
Sc$_2$CX (X=N$_2$, ON, O$_2$) MXenes as a Promising Anode Material:A First-Principles Study]{
Sc$_2$CX (X=N$_2$, ON, O$_2$) MXenes as a Promising Anode Material: A First-Principles Study}
\author{S.~\"Ozcan$^{1}$ }
 \email{sozkaya@aksaray.edu.tr}
 \affiliation{Department of Physics, Aksaray University, 68100 Aksaray, Turkey}
\author{B. Biel$^{1}$}%
 
\affiliation{ 
 Department of Atomic, Molecular and Nuclear Physics \& Instituto Carlos I de Física
Teórica y Computacional, Faculty of Science, Campus de Fuente Nueva, University of Granada, 18071 Granada, Spain}%

\date{\today}

\begin{abstract}
MXenes' tunable properties make them excellent candidates for many applications in future nanoelectronics. In this work, we explore the suitability of Sc$_2$CX (X=N$_2$, ON, O$_2$) MXenes to act as the active anode materials in Na-ion based batteries (NIBs) by means of \textsl{ab initio} simulations. After analyzing the structural and elastic properties of all the possible models to evaluate the energetically favorable N and O functionalization sites, our calculations show that both Sc$_2$CON and Sc$_2$CN$_2$ present a clear metallic character, making them potential candidates as anode materials. The investigation of the most relevant features for anode performance, such as the adsorption and diffusion of Na atoms, the intrinsic capacity, the open circuit voltage, and the storage capacity show that both systems are serious alternatives to the most common 2D materials currently employed in alkali metal batteries. In particular, Sc$_2$CN$_2$ presents a better diffusion behavior thanks to the absence of Na clustering on its surface, with optimal diffusion barriers comparable to other 2D materials such as MoN$_2$, while the values of diffusion barriers for Sc$_2$CON are at least three times smaller than those found for other anode candidates. Similarly, while the capacity of Sc$_2$CON is close to the one reported for 2D Sc$_2$C, Sc$_2$CN$_2$ possesses a power density more than twice higher than the ones of 2D materials such as Sc$_2$C, graphite, and MoS$_2$. Our results thus confirm the urge for further experimental exploration of the MXene Sc$_2$CX (X=N$_2$, ON, O$_2$) family as anode material in NIBs. 
\end{abstract}
\maketitle

\section{\label{sec:int}Introduction}

The 21$^{st}$ century triggered a growing demand for high-performance rechargeable lithium-ion batteries (LIB) because of their excellent reversible capacity, high energy density, and long cycle life ~\cite{Idota-97,Eshetu-21,Yoo-08,Armand-05,Arico-05}.
Although LIBs have gained great success as a power source for efficient energy storage systems, their high cost, safety risks, and the low abundance of lithium in the earth’s crust ~\cite{Goodenough-10, Goodenough-13, Tarascon-01} make it mandatory to search for batteries based on an alternative ion. Na, for instance, presents similar chemical properties to lithium at a much lower cost, thanks to the abundance of its natural reserves. For this reason, Na-ion batteries (NIBs) have been studied in recent years as an adequate candidate to replace LIBs in large-scale energy storage applications in the future~\cite{Wang-16, Hwang-17, Dai-17}. However, the development of NIBs is facing one crucial problem: finding anode materials with similarly excellent electrochemical performance to that currently achieved for LIBs. For LIBs, despite their relatively low capacity (372 mAhg$^{-1}$), graphite is almost invariably used as an anode material ~\cite{Dahn-95,Yoshio-03}. However, due to the larger radius of the sodium ions compared to that of lithium ones, graphite could lead to even lower capacities~\cite{He-14} when employed as an anode for NIBs. Several other carbon-based materials, such as graphene ~\cite{Wang-09, Buldum-13, Lee-21} or carbon nanotubes ~\cite{Zhou-00}, metallic VS$_2$ monolayer,~\cite{Jing-13}, have also been widely studied both theoretically and experimentally for LIBs. Despite several materials having been identified as potential candidates for electrodes in LIBs, so far, no ideal candidate for NIBs has been found.\\

In this context, novel two-dimensional (2D) materials with intriguing mechanical, electronic, and transport properties are being investigated as promising candidates to act as NIBs' anodes ~\cite{Shi-17}. Among them, the MXenes family is receiving more and more attention due to their excellent mechanical properties and good electrical conductivity~\cite{Aslam-20}, which makes them excellent candidates to act as anode materials for alkali metal ion batteries. For instance, Naguib $et \ al$. experimentally demonstrated that Ti$_2$C sheets exhibit a stable capacity of 225 mAhg$^{-1}$ in LIBs ~\cite{Naguib-12}. Ti$_3$C$_2$ MXene was also reported as an anode material with a storage capacity of 146 mAhg$^{-1}$) ~\cite{Xie14}. Additionally, 2D Nb$_2$C and V$_2$C based materials, synthesized experimentally by selective etching, show reversible capacities of 170 and 269 mAhg$^{-1}$ ~\cite{Naguib-13}, respectively. Interestingly, recent experiments and theoretical calculations have reported that the energy storage capacities of MXenes strongly depend on the functional groups terminating the otherwise bare surfaces of the MXenes  ~\cite{Peng-14,Xie-14, Yu-16, Zhao-17, Khazaei-17}. For example, while the hydroxyl (OH) and fluorine (F) surface groups reduce the storage capacity of metal ions ~\cite{Tang-12, Hu-14, Zhang-20}, some oxygen (O) functionalized MXenes (f-MXenes) have been found to be very suitable as anode materials in batteries ~\cite{Zhenhua-22}. Therefore, systematically studying the storage capacity changes associated with all the potential MXenes' surface functional groups is a pivotal step to optimizing the capacity levels of these materials. In particular, in order to improve the electrochemical performance of alkali metal ion batteries functional groups based on selenium (Se) ~\cite{Li19}, sulfur (S) ~\cite{Zhu-16}, phosphorus (P) ~\cite{Zhu-17}, silicon (Si) ~\cite{Zhu-17}, carbon (C) ~\cite{Xiao-19}, and nitrogen (N) ~\cite{Shao-17} have been also studied theoretically. Following theoretical predictions, sulfur, selenium, and tellurium functionalized MXenes have been recently synthesized experimentally ~\cite{Kamysbayev-20}.

On the other hand, the mirror asymmetry of Janus materials makes them particularly suitable to serve as anode materials, since the asymmetry causes an intrinsic out-of-plane polarized electric field across the material ~\cite{Li18}. For this reason, after the successful synthesis of the Janus MoSSe by Zhang $et \ al$. ~\cite{Zhang-17}, it was theoretically suggested that the use of MoSSe as an anode electrode in LIBs would lead to a small diffusion barrier (0.24 eV) and a high capacity (776.5 mAhg$^{-1}$) ~\cite{Wang-19, Shang-18}. Later on, MoSSe, TiSSe, and VSSe Janus materials were also proposed for both NIBs and potassium-based batteries (KIBs) ~\cite{Wang-19,Xiong-20}. 

In this context, Janus MXenes combine the tunability of functionalized MXenes to be optimized for the desired application with the asymmetry of Janus materials and are thus an optimal candidate for batteries based on alternative alkali metals. Recently, Edirisuriya $et \ al$. investigated the Ti$_2$CSO and Ti$_2$CSSe Janus MXene structures for electrodes in Li- and Mg-based batteries ~\cite{Edirisuriya-21}, finding that the Ti$_2$CSO monolayer exhibits a high gravimetric capacity (524.54mAhg$^{-1}$). However, to our knowledge, the suitability of Janus MXenes to act as electrodes in NIBs has been only barely explored. In particular, Sc$_2$C has the highest surface area to weight ratio of all MXenes and a high Li (462 mAhg$^{-1}$)/ Na (362 mAhg$^{-1}$) storage capacity ~\cite{XingshuaiLv-17}. It is, therefore, only natural to ascertain the possibility of using both Sc$_2$CN$_2$ and Sc$_2$CON monolayers as electrodes for the next generation of NIBs. The goal of our work is hence to explore, by calculations based on Density Functional Theory (DFT), the potential of the 
Sc$_2$CX (X=N$_2$, ON, O$_2$) MXene to serve as anode material for NIBs.  


\section{\label{sec:level2}Method}

All the calculations were carried out with the Vienna Ab Initio Simulation Package (VASP)~\cite{vasp1,htt,Kre-99,Kres-96}, based on
DFT. In this method, the Kohn--Sham single particle functions were
expanded based on plane waves up to a cut-off energy of 500 eV. The
electron--ion interactions were described by using the projector
augmented-wave (PAW) method~\cite{Kre-99,Bl-94}. For electron
exchange and correlation terms, Perdew--Zunger-type
functional~\cite{Per-81,Per-92} was used within the generalized
gradient approximation (GGA)~\cite{Bl-94} in its PBE parametrization ~\cite{Perdew-1996}. Given the well-known underestimation of the band gap values by this functional, we have refined our calculations by using the Heyd–Scuseria–Ernzerhof (HSE06) functional, since it provides more accurate
band gaps and band edge positions than the PBE   ~\cite{Heyd-2003, Heyd-2006, Krukau-2006}. 
The self-consistent solutions were obtained by employing the
($11\times11\times1$) Monkhorst--Pack~\cite{Mon-76} grid of k-points for
the integration over the Brillouin zone for $2\times2$ supercell ~\cite{Zhenhua-22, Zhu-16, Chen21, Hu16, Lia20}. To prevent spurious interaction between isolated layers, a vacuum layer of at least 16 {\AA} was included
along the normal direction to the surface for both sides of the slab. The elastic tensor for each system was derived from the stress-strain approach ~\cite{Le-02}.

\begin{figure}[h] \centering
	\includegraphics*[width=8cm,clip=true]{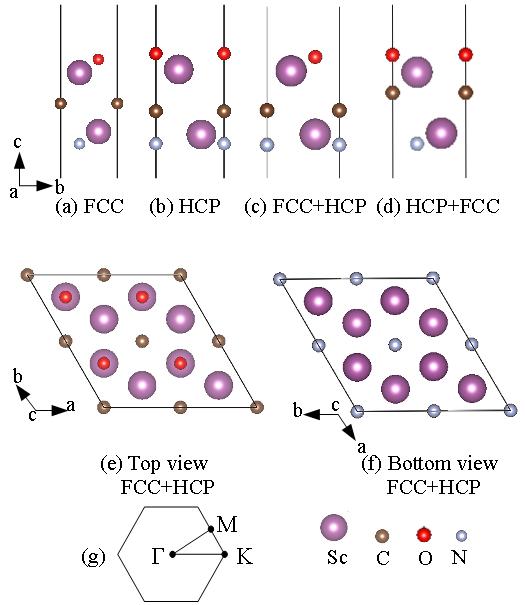}
	\caption{(a-d) Side views of the four possible structures for Janus Sc$_2$CON. (e-f) Top and bottom views of the FCC+HCP structure of bare Sc$_2$CON. Since the only change concerning Sc$_2$CN$_2$ (Sc$_2$CO$_2$) is the substitution of the O (N) atoms on the top (bottom) layer for N (O) atoms, here we only present the Sc2CON structure. Sc atoms are shown in purple, C atoms in brown, O atoms in red, and N atoms in blue. The Brillouin zone of Sc$_2$CN$_2$, Sc$_2$CO$_2$, and Sc$_2$CON, showing the high symmetry points.}
	\label{models}
\end{figure}

\section{\label{sec:results}Results and Discussions}
\subsection{Structural and mechanical properties}
We start by looking for the energetically favorable configurations of the optimized Sc$_2$C for all the possible O or N adsorption geometries. Depending on the relative positions of the termination groups attached to the transition metal atoms, four different configurations are possible for the chemical termination of Sc$_2$CON, presented in Fig. 1 as FCC, HCP, FCC+HCP, and HCP+FCC. To fully understand the impact of the asymmetric surface termination in the Sc$_2$CON structure, we have also analyzed the atomic and electronic structures of both Sc$_2$CN$_2$ and Sc$_2$CO$_2$. 
For the sake of clarity, we only show the Sc$_2$CON structure here (please see Fig.S1 for Sc$_2$CO$_2$ and Sc$_2$CN$_2$), since the only change concerning the Sc$_2$CN$_2$ (Sc$_2$CO$_2$) is the substitution of one of the O (N) atoms on the top (bottom) layer for an N (O) atom. 
In the FCC model, both O (or N) and N atoms align with the Sc atoms, while in the HCP model these atoms are positioned above and below the C atoms. In the FCC+HCP model, the O atoms occupy the same place they do in the FCC configuration, whereas the N atoms are placed as in the HCP model. On the contrary, in the HCP+FCC model, the N atoms are the ones positioned as in the FCC structure, while the O atoms are placed as in the HCP geometry. After full optimization of the Sc$_2$CON, Sc$_2$CO$_2$, and Sc$_2$CN$_2$ MXenes, our results for the lattice parameters, relative total energies with respect to the  most stable structure, formation energies concerning the structure with the lowest energy, and band gaps are presented in Table 1. The most energetically stable model for all the considered MXenes is the FCC+HCP, while the FCC structure exhibits the highest relative energy and it is thus the less stable. In agreement with a previous study ~\cite{Li-16}, Sc$_2$CO$_2$ also has FCC+HCP as the lowest energy structure. We can evaluate the formation energies of each configuration by calculating their formation energies as per Equation 1:\\
\begin{table*}
	\centering 
	\caption{Lattice parameters ($a$ in {\AA}), relative total energies ($\Delta$E in eV) (please see Table S1. for the total energies), formation energies ($E_f$ in eV) with respect to the lowest energy structure, and band gaps ($E_g$ in eV) for the Sc$_2$CO$_2$, Sc$_2$CN$_2$, and Sc$_2$CON MXenes. ID indicates the indirect band gap semiconductor character. Note that the FCC+HCP and the HCP+FCC geometries are the same structural models in Sc$_2$CO$_2$ and Sc$_2$CN$_2$ since both the top and bottom surfaces are equally terminated.}
	
	\centering \vspace{2mm} \centering
	\footnotesize\setlength{\tabcolsep}{8pt}
	\begin{tabular}{c| c c| c c| c c| c c| c c }
		\hline
		& \multicolumn{2}{c}{FCC} & \multicolumn{2}{c}{HCP}& \multicolumn{2}{c}{FCC+HCP}& \multicolumn{2}{c}{HCP+FCC} \\ 
		
		 & ~$a~$ &  $\Delta$E &~$ a ~$&  $\Delta$E &~$ a~$ &  $\Delta$E &~$ a~$ &  $\Delta$E &$E_f$ & $E_g$ \\
		\hline
		
	 	 Sc$_2$CO$_2$	   & 3.22 & 0.66	& 3.37	& 0.14	& 3.42 &	0.00 &	-- &	-- & -10.40 &  	2.84 (ID) \\[0ex]
	 	 
	 	 Sc$_2$CN$_2$	    & 3.58 & 3.45	& 3.58	& 0.95	& 3.42 &	0.00 &	-- &	-- & -8.40 &	Metal  \\[0ex]
			
	 Sc$_2$CON	   & 3.36 & 2.88	& 3.45	& 0.56 &	3.48 &	0.00 &	3.51 & 2.04 & -9.85&	Metal  \\[0ex] 
		\hline
	\end{tabular}
\end{table*}

\begin{equation}
\begin{gathered}
$$$ $E_f$ = $E_{tot}$(Sc$_2$CON) – $E_{tot}$(Sc$_2$C) – $E_{tot}$(O) - E$_{tot}$(N), $$$ \\
$$$ $E_f$ = $E_{tot}$(Sc$_2$CY$_2$) – $E_{tot}$(Sc$_2$C) – 2$E_{tot}$(Y) , $$$
\end{gathered}
\end{equation}

where $E_{tot}$(Sc$_2$CON), $E_{tot}$(Sc$_2$CY$_2$) (Y=O or N), $E_{tot}$(Sc$_2$C), $E_{tot}$(O), E$_{tot}$(N), E$_{tot}$(Y) are the total energy of the surface terminated Sc$_2$CON, Sc$_2$CO$_2$ or Sc$_2$CN$_2$, the total energy of pristine Sc$_2$C, and the total energy of (1/2) (O$_2$ in the gas
phase), N (stable bulk form), (1/2) (O$_2$ in the gas
phase) or N (stable bulk form), respectively.





The negative formation energies ($E_f$) of all the MXenes considered in this work show that the synthesis of these MXenes is highly possible. This is further confirmed by the analysis of the Born mechanical stability criteria that we present below. From now on, we will concentrate on the most stable geometry, the FCC+HCP.

The lattice constant ($a$) of the Sc$_2$CON MXene (for the most stable model) is 3.48 {\AA}, with a distance between the O and the N atom layer of 3.66 {\AA}. The similar size of O and N atoms results in the same lattice constant of 3.42 {\AA} for both Sc$_2$CO$_2$ and Sc$_2$CN$_2$, slightly shorter than the one obtained for Sc$_2$CON, and with a distance between the O-O and the N-N atom layers of 3.86 {\AA}.
\begin{table}[b]
	\caption{Elastic constants ($C_{ij}$ in N/m), Young’s moduli (Y in N/m), 
		 and Poisson’s ratios ($\nu$) for the Sc$_2$CO$_2$, Sc$_2$CN$_2$, and Sc$_2$CON MXenes.}
	\centering \vspace{2mm} \centering
	\begin{tabular}{l|cccccc}
		\hline &&&&&\\[-3ex]
		~Material~ & ~$C_{11}$~ & ~$C_{12}$~&~$C_{66}$~&$Y$~  & ~$\nu$~\\[0.5ex]
	\hline
		~Sc$_2$CO$_2$~ &168.798&	50.552&	59.123&	153.658&	0.299~ \\[0.5ex]
	
		~Sc$_2$CN$_2$~ &203.749&	95.810&	53.969&	158.695&	0.470~ \\[0.5ex]
		
		~Sc$_2$CON~ &157.733&	53.363&	52.185&	139.680&	0.338~ \\[0.5ex]
	
		\hline
	\end{tabular}
	\label{table-2}
\end{table}

Mechanical properties also play a critical role in the designing of 2D nanodevices. For this reason, we have investigated the elastic properties of this MXene family. Our results are shown in Table 2. We have verified that the Sc$_2$CO$_2$, Sc$_2$CN$_2$, Sc$_2$CON Mxenes satisfy the Born mechanical stability criteria ~\cite{Le-02},

\begin{equation}
\label{eq2}
	\ C_{11}>0, \;\ C_{66}>0, \;\
	\ 2*C_{66}=C_{11}-C_{12}, \;\  and \ C_{11}> |C_{12}| 
\end{equation}

further implying that they are mechanically stable. According to the calculated elastic constant, we calculated the Young’s modulus Y and the Poisson’s ratio $\nu$ as below ~\cite{Thomas-2018,Wei-2009,Andrew-2012}

\begin{equation}
\label{eq3}
\begin{split}
 	\ Y=C_{11}-C_{12}*(\nu), \;\ \nu =C_{12}/C_{11}
	\\
\end{split}
\end{equation}

\begin{figure*} 
\includegraphics*[width=\textwidth]{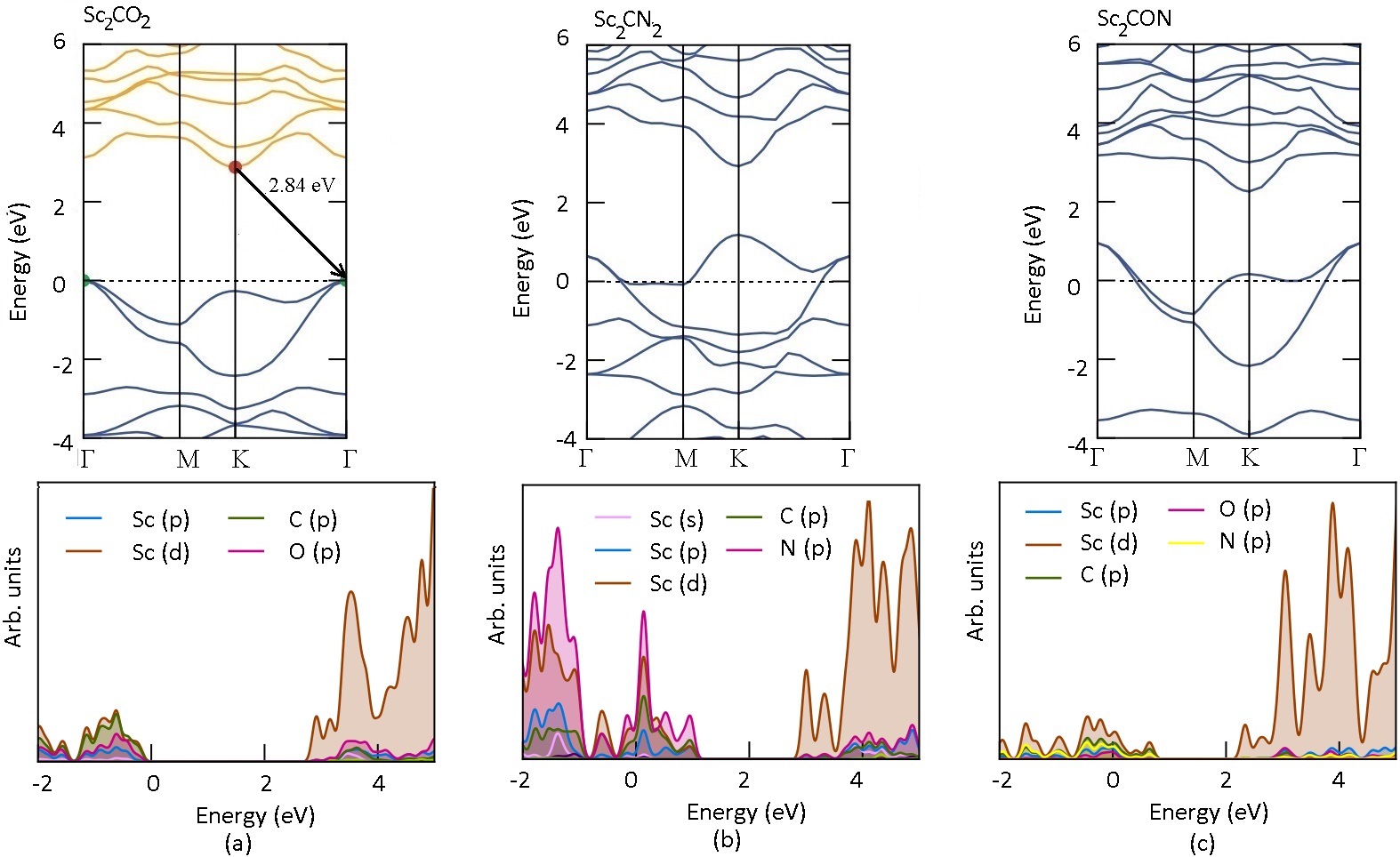}
	\caption{Band structure (top panel) and PDOS (bottom) of (a) Sc$_2$CO$_2$ (b) Sc$_2$CN$_2$ (c) Sc$_2$CON. The Fermi energy is set at zero. For Sc$_2$CO$_2$, the valence band (blue lines) maximum and conduction band (orange lines) minimum are highlighted with green and red points, respectively. The dashed line denotes the Fermi level.}
	\label{models}
\end{figure*}

The Young's modulus of Sc$_2$CX (X=ON, N$_2$, O$_2$) decreases as Y(Sc$_2$CN$_2$) $>$ Y(Sc$_2$CO$_2$) $>$ Y(Sc$_2$CON). All of the three materials studied here present a similar value of Y (139 N/m - 159 N/m), a much larger value than the one reported for other 2D materials such as monolayer WS$_2$ (106.4 N/m), silicene (62 N/m)~\cite{Mortazavi-2017} and germanene (44 N/m)~\cite{Mortazavi-2017}. This means that the Sc$_2$CX (X=ON, N$_2$, O$_2$) would be more suitable for applications where strain is an issue.

\subsection{Electronic properties}

A critical feature for any material candidate to be used as a battery anode is its electrical conductivity. To evaluate the electronic properties of the Sc$_2$CX (X=ON, N$_2$, O$_2$) MXenes, we calculated their band structures, shown in Fig. 2 (top panel). 
Interestingly, while Sc$_2$CO$_2$ is a semiconductor with an indirect band gap (E$_g$) of 2.84 eV, both Sc$_2$CN$_2$ and Sc$_2$CON show a metallic behavior, which implies their great potential to enhance anode performance in energy-storage applications. The analysis of the partial density of states (DOS) presented in the bottom panel of Fig. 2 allows us to gain further insight into the electronic structure of these materials, revealing that the metallicity mainly originates from the $p$ orbital of the N atom for Sc$_2$CN$_2$ and the $d$ orbital of the Sc atoms in Sc$_2$CON. For Sc$_2$CO$_2$, the states above the Fermi energy level (EF) come from the $d$ orbital of the Sc atoms, while the states below EF are occupied by $d$ orbital of the Sc atoms and  $p$ orbital of the O atoms (in the deeper levels).

\subsection{Adsorption and diffusion of Na atoms} 

Two more essential parameters to characterize a material's suitability to act as an electrode in electronic devices are the adsorption and diffusion energies of certain chemical species, such as Na, onto their surface. The adsorption of ions on the surface of the electrode is indeed related to the charge storage capacity of this material, while the diffusion gives us information on the electronic mobility and the charge/discharge rate of the potential electrode. To analyze the potential of the metallic Sc$_2$C MXenes for this purpose it is necessary, firstly, to calculate the adsorption energy of Na on  one surface of Sc$_2$CN$_2$ and Sc$_2$CON. The maximum number of Na atoms that can be adsorbed in one surface of the monolayer, for a $2\times2$ supercell, is four. Fig. 3 displays the 4 possible adsorption configurations, T1 - T4. The top and bottom surfaces of the FCC+HCP structure are not symmetric, which means that the Na atoms will occupy different positions on each of the surfaces. For the T1 (T3) configuration, the Na atoms will sit on the top (bottom) of the top layer Sc atoms (see Fig.3). For the T2 and T4 configurations, the Na atoms will sit on top of the C atoms of the middle layer and below the Sc atoms of the bottom layer, respectively. This means that one Na atom can be adsorbed on Sc$_2$CN$_2$ and Sc$_2$CON with 4 different possible adsorption sites. 
 
\begin{figure}[h] \centering
	\includegraphics*[width=8.5 cm,clip=true]{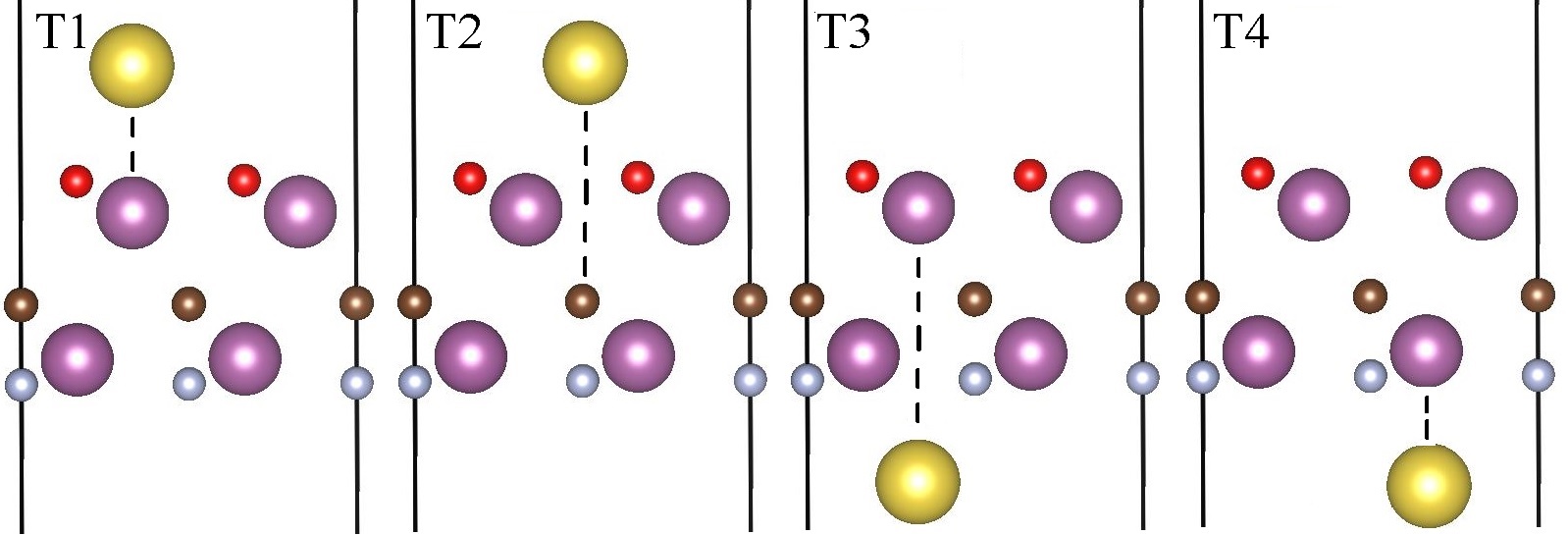}
	\caption{Side views of the possible structures for Na single-ion adsorption on a $2\times2$ supercell in Sc$_2$CON and Sc$_2$CN$_2$ MXenes. Since both MXenes share the same FCC+HCP structural model we only show the case of Sc$_2$CON. Sc atoms are shown in purple, C atoms in brown, N atoms at the bottom layer in blue, N and/or O atoms at the top layer in red, and Na atoms in yellow. }
	\label{models}
\end{figure}

The adsorption energy for a single Na adatom adsorption can be calculated by the equation below:

\small
\begin{equation}
\begin{gathered}
$$$$E_{ads}$ = ($E_{tot}$(Sc$_2$CX)+$x$Na)- $E_{tot}$(Sc$_2$CX) - $x$ $E_{tot}$(Na))/$x$$$$
\end{gathered}
\end{equation}

where $E_{tot}$(Sc$_2$CX)+$x$Na is the total energy of the whole system, $E_{tot}$(Sc$_2$CX) is the energy of the pure Sc$_2$CX (X=ON, N$_2$), $E_{tot}$(Na) represents the total energy of a single Na atom in the bulk, and $x$ is the number of Na ions on both surface layers according to the available sites on each surface.

\begin{table}[b]
	\caption{Adsorption energy ($E_{ads}$) in eV for the possible adsorption sites of Sc$_2$CX (X=ON, N$_2$).}
	\centering \vspace{2mm} \centering
		\begin{tabular}{l|ccccc}
		\hline &&&&&\\[-2ex]
		~Material~ & ~$E_{ads}$(T1)~ & ~$E_{ads}$(T2)~&~$E_{ads}$(T3)~&$E_{ads}$(T4)~\\[0.5ex]
	
		\hline &&&&&\\[-2ex]
	
		~Sc$_2$CN$_2$~ &-1.40&-1.85&	-0.05&	-0.69 ~ \\[0.5ex]
		
		~Sc$_2$CON~ & -0.79 &	-1.08 &	-0.36 &	-0.32 ~ \\[0.5ex]
	
		\hline
	\end{tabular}
	\label{table-3}
\end{table}

The calculated adsorption energies for the Na atom on each configuration are tabulated in Table 3. All the adsorption energies have negative values, indicating that the Na atom could be adsorbed at any of the locations and the Na/MXene system is stable. In our convention, more negative adsorption energy denotes a more favorable exothermic reaction between the Na atom and Sc$_2$CN$_2$ or Sc$_2$CON. The most favorable position for Na atom adsorption is the T2 site for both the Sc$_2$CON and the Sc$_2$CN$_2$. 

We must note, however, that when an undesirable metal clustering occurs (such as it has been found, for instance, for pure graphene), the performance of a battery application becomes limited by the weak binding of the metal on the material ~\cite{Karmakar-16, Lee-12, Fan-13}. The adsorption energy of Na on Sc$_2$CN$_2$ (-1.85 eV) is lower than the cohesive energy of Na (-1.13 eV) ~\cite{Aver-72}. However, the value of $E_{ads}$ for Sc$_2$CON (-1.08 eV) is very similar to the Na cohesive energy. This means that Sc$_2$CN$_2$ would be less prompt to cluster and would thus potentially behave better as anode material for Na ion batteries. Similar behavior was found in other theoretical studies for the Na atom on the surface of WTe$_2$ ~\cite{Sarvazad-19} and the two-dimensional sheet of germanium selenide ~\cite{Sannyal-18}.

\begin{figure*} 
\includegraphics*[width=\textwidth]{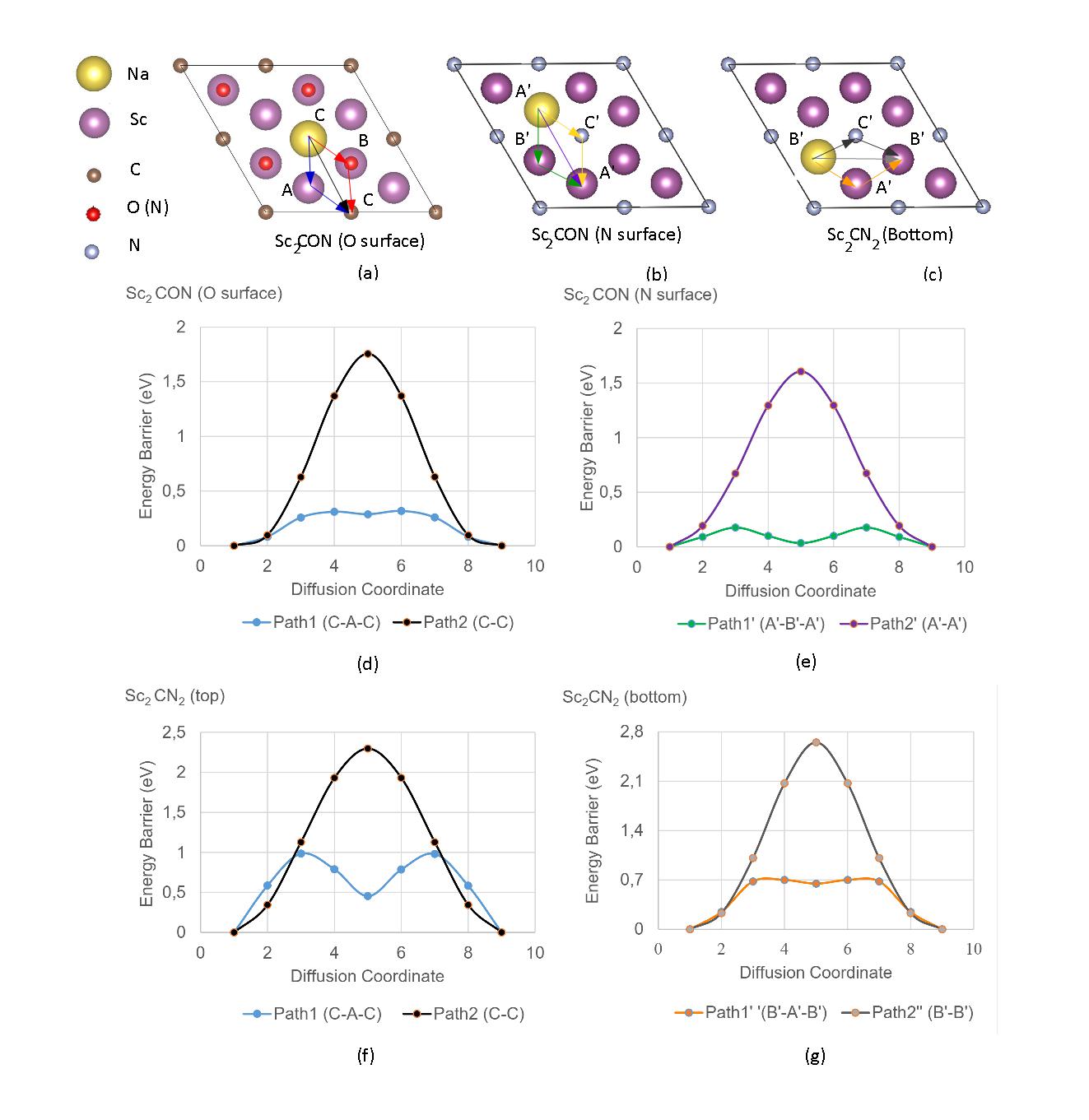}
	\caption{Top panel: Diffusion pathways for (a) Sc$_2$CON (top surface) (b) Sc$_2$CON (bottom surface) and (c) Sc$_2$CN$_2$ (N-terminated surface.) Please note that the top surface of Sc$_2$CN$_2$ coincides with the O-terminated surface of Sc$_2$CON, so only one geometry is shown here. Besides, although both the bottom surface of Sc$_2$CN$_2$ and the N-terminated surface of Sc$_2$CON are the same, the Na adsorption sites differ, and so do the pathways for each material. Middle panel: Calculated diffusion barriers for Sc$_2$CN$_2$ for the top (d) and bottom (e) surfaces.
	Bottom panel: Calculated diffusion barriers for Sc$_2$CON for the top (f) and bottom (g) surfaces.}

	\label{models}
\end{figure*}

Furthermore, one of the most important characteristics of an electrode material for rechargeable batteries is the easy flow of ions inside the anode material, which is strongly related to
the performance of the battery. We have calculated the diffusion barriers using the nudged elastic band (NEB) method to provide more insight into the diffusion properties of Na on the surface of the Sc$_2$CN$_2$ and Sc$_2$CON monolayers. There are three possible paths between the nearest neighboring favorable adsorption sites: Path 1 (C → A → C), Path 2 (C → C), and Path 3 (C → B → C), as shown with blue, black, and red arrows in the top panel of Fig. 4a, respectively. Since the functional groups, O and N are located higher on the surface than the Sc and C atoms, the Na atom has to overcome a strong interaction barrier when it climbs over the functional group atoms in Path 3. For this reason, the diffusion barrier of Path 3 will be extremely high compared to those of the other two paths, and we have calculated only the diffusion barriers for Path 1 and Path 2. 

We found that, for both systems, Path 1 gives the lowest diffusion barrier, so we will only focus on this potential trajectory. The calculated diffusion barriers for the Na atom along Path 1 (C→A→C)
are 0.32 eV and 0.98 eV on  Sc$_2$CON (O surface) and Sc$_2$CN$_2$ (top surface ), respectively (see Fig.4d and Fig.4f, blue lines). 

Since Sc$_2$CN$_2$ and Sc$_2$CON present a different symmetry at the top and bottom surfaces, we need to calculate $E_{diff}$ also for the bottom surfaces. We will name the trajectories of the bottom surface of Sc$_2$CON as Path 1' (A'→ B'→ A') and Path 2' (A'→ A') (green and purple arrows in Fig.4b, respectively.) 
The possible diffusion pathways for Sc$_2$CN$_2$ are Path 1'' (B'→ A'→ B') and Path 2'' (B'→ B'), as shown in Fig. 4c with the orange and gray arrows, respectively. Again, the lowest barriers are found for Path 1' in Sc$_2$CON and Path 1'' in Sc$_2$CN$_2$, with values of 0.18 eV and 0.72 eV, respectively. The diffusion curves are depicted in Figures 4e and 4g. 


\begin{figure}[h] \centering
	\includegraphics*[width=6cm,clip=true]{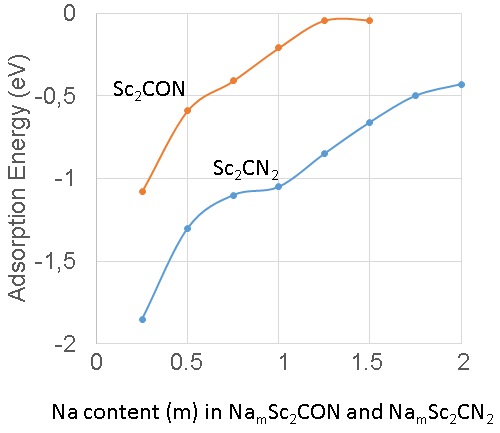}
	\caption{Variation in the adsorption energy with increasing Na content on the Sc$_2$CN$_2$ (blue line) and Sc$_2$CON (orange line) monolayers.} 
	\label{models}
\end{figure}

The diffusion barrier value for the bottom surface of Sc$_2$CN$_2$ is 0.72 eV, comparable to those found for Na and Li diffusion on MoN$_2$, 0.57 eV and 0.78 eV, respectively \cite{Zha-16}, and the one found for Na diffusion on Ti$_3$C$_2$S$_2$, 0.6 eV \cite{Men-18} and TiO$_2$-based polymorphs, 0.65 eV \cite{Wag-01}. 
However, the value found for the N-terminated surface of Sc$_2$CON, 0.18 eV, is much smaller than those found for the diffusion of alkali metals on other 2D energy storage materials, such as 0.33 eV for Li diffusion on graphene \cite{Pol-10} and 0.25 eV for Li diffusion on MoS$_2$ \cite{Dav-14}. On the other hand, the diffusion barriers of Na are about an order of magnitude larger than in pure  Sc$_2$C (0.012 eV  \cite{XingshuaiLv-17}). A similar result has also been found for 2D Ti$_2$CT$_2$ (T=C, O and S) monolayer MXenes as anodes for Na-ion batteries by Xiao $et \ al$. \cite{Xiao-19}. In those works, it was reported that the diffusion energy barriers follow the order of bare surface < O-terminated surface < S-terminated surface < C-terminated surface. These results reveal that while the sodium ions could easily migrate on the pure MXene monolayer, the functional groups tend to impede the sodium-ions diffusion \cite{Xiao-19}.


\subsection{Intrinsic capacity}

As we have already mentioned, the maximum Na capacity on Sc$_2$CN$_2$ and Sc$_2$CON was determined using a $2\times2\times1$ supercell. Since the maximum number of Na atoms adsorbed on the top and bottom surfaces is 8, the concentration $n$ of Na atoms is within the range $n$=0.25-2. The changing trend of $E_{ads}$ with the concentration of Na is displayed in Fig. 5. The adsorption energy increases with the increase in $n$, which can be explained as follows: as the Na atom gives electrons to the Sc$_2$CN$_2$ (or Sc$_2$CON) monolayer, it gets positively charged. The increasing number of Na$^+$ ions hence generates a strong repulsive force between the Na$^+$–Na$^+$ ions that overcomes the electrostatic force between Na$^+$ and N (or O). Thus, the Na$^+$ ion will tend to separate from the Sc$_2$CN$_2$ (or Sc$_2$CON) surface, increasing the adsorption energy.

For the case of Sc$_2$CON, we stopped the calculations when the adsorption energy turned positive at a concentration of $n$=1.5. Thus, the maximum capability of Na storage of the Sc$_2$CON monolayer is 6 Na atoms (Na$_{1.5}$Sc$_2$CON). In contrast, a negative adsorption energy of -0.43 eV/atom persisted at $n$=2 for Sc$_2$CN$_2$. This value is much larger than those reported for other typical electrode materials. For example, the adsorption energy of Na on Sc$_2$C (-0.32 eV/atom) ~\cite{XingshuaiLv-17}, GeS (-0.02 eV/atom) ~\cite{Hu-15}, and Ca$_2$N (-0.003 eV/atom)~\cite{Hu-15}. Therefore, 
the greater adsorption capacity of Sc$_2$CN$_2$ makes it much more suitable to become the active anode material in Na-based batteries than other 2D materials.

To evaluate the feasibility of multilayer adsorption, we calculated the average adsorption energies as the number of Na atoms was increased further, with the average adsorption energy defined by\\

\begin{align}
 E_{av}=(E_{tot}(Na_{(8+m)}Sc_8C_4N_8)-E_{tot}(Na_8Sc_8C_4N_8)-m_{Na})/m 
\end{align}

($m$=1,2,...,8) where $E_{Na}$ is the cohesive energy of the metal, $E_{tot}(Na_8Sc_8C_4N_8)$ is the total energy of the Sc$_2$CN$_2$ supercell with one Na adlayer on both sides (Na$_8$Sc$_8$C$_4$N$_8$), and $E_{tot}(Na_{(8+m)}Sc_8C_4N_8)$ represents the total energy of Na$_{(8+m)}$Sc$_8$C$_4$N$_8$ with $m$ extra Na atoms added to the Na$_8$Sc$_8$C$_4$N$_8$ system.

After the adsorption sites of both surfaces of Sc$_2$CN$_2$ were completely saturated with Na atoms, we proceeded to add one additional Na atom on the second layer similarly as it was done in the first step to studying the adsorption of the first Na layer. We found that the most stable configuration is the T3 (in Fig.3), where the new Na atom is aligned under the Sc of the first layer, as shown in Fig.6. After relaxation, the maximum number of Na storage on the Sc$_2$CN$_2$ is 12 (8 in the first Na layer + 4 in the second Na layer), and the corresponding chemical stoichiometry is Na$_3$Sc$_2$CN$_2$. 

\begin{figure}[h] \centering
	\includegraphics*[width=2cm,clip=true]{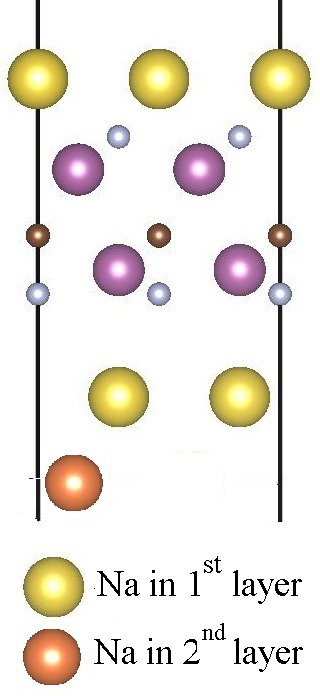}
	\caption{The adsorption site of second layer Na atom on the Sc$_2$CN$_2$. Sc atoms are shown in purple, C atoms in brown, N atoms in blue, Na first layer in yellow, and Na second layer in orange.} 
	\label{models}
\end{figure}


\subsection{Theoretical open circuit voltage and storage capacity} 

Finally, to determine the energy storage performance, the Na theoretical storage capacity ($C$) and the Open Circuit Voltage (OCV) were explored. The theoretical storage capacity was obtained according to Equation (6)~\cite{Yu-16},

\begin{equation}
\begin{gathered}
$$$ $C_M$=$(xzF)/(3.6M_{Sc_2CX})$ 
$$$
\end{gathered}
\end{equation}

where $z$ is the ionic valency of sodium ($z$ = 1), $x$ is the maximum Na content, $F$ is the Faraday constant (F = 96485.3C/mol), and $M_{Sc_2CX}$ 
is the molar mass of the host material Sc$_2$CX (X=ON or N$_2$) in g/mol. Theoretical capacities of 618 mAhg$^{-1}$  and 304 mAhg$^{-1}$ were obtained for Na adsorption on Sc$_2$CN$_2$ and Sc$_2$CON, respectively. While the capacity of Sc$_2$CON is close to the case of 362 mAhg$^{-1}$ reported for  Sc$_2$C monolayer ~\cite{XingshuaiLv-17}, the Sc$_2$CN$_2$ monolayer possesses a power density higher than the ones of Sc$_2$C monolayer and other 2D materials, such as graphite (372 mAhg$^{-1}$) ~\cite{Zhu-10} and MoS$_2$ (146 mAh·g$^{-1}$ ) ~\cite{Mortazavi-14}. This again corroborates the identity of the Sc$_2$CN$_2$ monolayer to be employed in alkali metal batteries.

Finally, we checked the OCV for Na atoms on both the Sc$_2$CN$_2$ and Sc$_2$CON monolayers, which is defined by  ~\cite{Li-19}: 

\begin{equation}
\begin{gathered}
$OCV$ \approx $(E$_{tot}$(Sc$_2$CX)+xE$_{tot}$(Na)- E(Sc$_2$CX+xNa))/x$ 
\end{gathered}
\end{equation}

where $E_{tot}(Sc_2CX)$, $E_{tot}(Sc_2CX+xNa)$, and $E_{tot}$(Na) represent the total energy of the bare Sc$_2$CX (X=ON or N$_2$), the Sc$_2$CX with x adsorbed Na cations, and a single bulk Na atom, respectively. When the adsorption energy turned to be positive, the OCV calculation was stopped. To prevent the dendrite formation of alkali metals and maximize the power density,  it’s accepted that the average OCV value must be within the range of 0-1.0 V for an anode ~\cite{Wang-20, Jing-19, Eames-14}. The OCV has been estimated for all the Na ion concentrations studied here on both the Sc$_2$CN$_2$ and Sc$_2$CON monolayers, as depicted in Fig. 7. The height of the blue (orange) column defines the value of the OCV. The average OCV values of Sc$_2$CN$_2$ and Sc$_2$CON, 0.76V and 0.40 V, respectively, meet the preferred voltage windows (between 0-1 V). Therefore, the suitability of Sc$_2$CN$_2$ and Sc$_2$CON to act as anode materials for NIBs is further confirmed.

\begin{figure}[h] \centering
	\includegraphics*[width=8.5cm,clip=true]{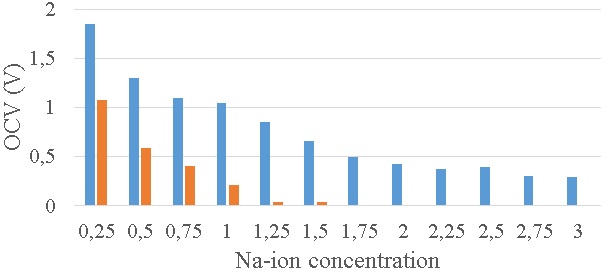}
	\caption{OCV profiles as a function of Na concentration on Sc$_2$CN$_2$ (blue columns) and Sc$_2$CON (orange columns).} 
	\label{models}
\end{figure}

\section{\label{sec:sum}Summary}

We have identified, by means of DFT-based simulations, the energetically favorable N and O functionalization sites for the Sc$_2$C MXene family. The investigation of their mechanical and electronic properties shows the potential of the metallic Sc$_2$CON and Sc$_2$CN$_2$ monolayers to act as anode materials in Na-ion batteries. To this aim have investigated the most relevant variables for anode performance, such as the adsorption and diffusion of Na atoms, the intrinsic capacity, the open circuit voltage, and the storage capacity, proving that both systems outperform the most common 2D materials currently employed in alkali metal batteries.\\

In particular, the analysis of their mechanical properties shows the excellent mechanical flexibility of functionalized MXenes under large strain, making them extremely suitable for applications where stress could be an issue. Interestingly, upon functionalization, Sc$_2$CO$_2$ become a semiconductor but both Sc$_2$CN$_2$ and Sc$_2$CON remain metallic, thus making them serious alternatives to the most common 2D materials currently employed in alkali metal batteries. For instance, Sc$_2$CN$_2$ presents a better diffusion behaviour thanks to to the absence of Na clustering on its surface, with optimal diffusion barriers comparable to other 2D materials such as MoN$_2$, while the values of diffusion barriers for Sc$_2$CON are at least three times smaller that those found for other anode candidates. Similarly, while the capacity of Sc$_2$CON is close to the one reported for 2D Sc$_2$C, Sc$_2$CN$_2$ possesses a power density more than twice higher than the ones of 2D materials such as Sc$_2$C and MoS$_2$.

Our results lay the ground for further experimental and theoretical work aimed to achieve the exploitation of these materials as 2D NIBs anodes.

\section*{Author Contributions}
Sibel Özcan: Conceptualization (lead); formal analysis (lead); writing – original draft (lead); writing – review and editing (equal). Blanca Biel: Conceptualization (supporting); writing – review and editing (equal); funding acquisition (lead).

\section*{Conflicts of interest}
There are no conflicts to declare.

\section*{Supplementary Material}
 See the supplementary material for side views of the four possible structures for Sc$_2$CO$_2$ and Sc$_2$CN$_2$, and the total energies of Sc$_2$CO$_2$, Sc$_2$CN$_2$, and Sc$_2$CON MXenes for the FCC, HCP, FCC+HCP, and HCP+FCC configurations.

\section*{Acknowledgments}

S.Ö. and B.B. kindly acknowledge financial support by the Junta de Andalucía under the Programa Operativo FEDER P18-FR-4834. B.B. also acknowledges financial support from AEI under project PID2021-125604NB-I00. The Albaicín supercomputer of the University of Granada and TUBITAK ULAKBIM, High Performance and Grid Computing Center (TRUBA resources) are also acknowledged for providing computational time and facilities.

\end{document}